\documentclass{lmcs}
\pdfoutput=1    

\usepackage{hyphenat}
\usepackage{mdwtab}
\usepackage{amsmath}
\usepackage{amssymb}
\usepackage{comment}
\usepackage[hyphens]{url}

\usepackage{hyperref}           

\usepackage{listings}

\lstdefinestyle{sOcaml}{language=[Objective]Caml,
  morekeywords={effect,let locus},
  literate={+}{{$+$}}1 {/}{{$/$}}1 
           {=}{{$=$}}1
           {>}{{$>$}}1 {<}{{$<$}}1
           {<>}{$\not=$}1
           {->}{{$\rightarrow$}}2 {>=}{{$\geq$}}2 {<-}{{$\leftarrow$}}2
           {<=}{{$\leq$}}2
           {==>}{{$\mapsto$}}2
           {|}{{$\mid$}}1
           {'a}{$\alpha$}1
           {'b}{$\beta$}1
           {'c}{$\gamma$}1
           {'e}{$\epsilon$}1
           {'k}{$\kappa$}1
           {'n}{$\nu$}1
           {'w}{$\omega$}1
           {'r}{$\rho$}1
           {'state}{$\sigma$}1
           {'w.}{$\forall\omega.\ $}2
           {'G}{\ensuremath{\Gamma}}1
           {vee}{\ensuremath{\vdash}}1
           {'l}{\ensuremath{\ell}}1
           {fn}{$\lambda$}1
           {e1}{e$_1$}1
           {e2}{e$_2$}1
           {e_m}{e$_m$}1
           {_1}{$_1$}1
           {_2}{$_2$}1
           {+\%}{{$\underline+$}}1
           {-\%}{{$\underline-$}}1
           {=\%}{{$\underline=$}}1
           { *\% }{{$\underline*$}}1
           {>\%}{{$\underline>$}}1
           {/\%}{{$\underline/$}}1
           {fnC}{$\underline\lambda$}1
           {@}{$\underline @$}1
           {@@}{@@}1
           {cint}{\underline{int}}4
           {cbool}{\underline{bool}}5
           {cif}{\underline{if}}3
           {:=}{\ensuremath{\mathrel{{:}{=}}}}2
           {ER[}{\ensuremath{\mathcal{E}_R\big[}}2
           {MR[}{\ensuremath{\mathcal{M}_R\big[}}2
           {MRR[}{\ensuremath{\mathcal{M}_R^R\big[}}2
           {MSR[}{\ensuremath{\mathcal{M}_R^S\big[}}2
           {]!}{\ensuremath{\big]}}1
           {...}{\ldots}2
           {empty}{\ensuremath\varnothing}1
           {\#\#\#}{{$\leadsto$}}3
}
\lstnewenvironment{code}[1][]{\lstset{basicstyle={\linespread{1.02}\sffamily\normalsize},#1}}{}

\lstMakeShortInline[breaklines=true]|

\newcommand{\quant}[2]{\mathopen{{\divide\thinmuskip3 \divide\medmuskip3 \divide\thickmuskip3 #1#2.}\mskip1mu}}
\newcommand{\lam}{\quant\lambda}
\newcommand\Nom{\mathcal{N}}
\newcommand\Loc{\mathcal{L}}

\newcommand{\BR}[1]{\big[\mathsf{#1}\big]} 
\newcommand{\TR}[1]{\ensuremath{\mathcal{T}_R\BR{#1}}}
\newcommand{\TS}[1]{\ensuremath{\mathcal{T}_S\BR{#1}}}
\newcommand{\TX}[1]{\ensuremath{\mathcal{T}_X\BR{#1}}}
\newcommand{\TUX}[1]{\ensuremath{\mathcal{T}^X\BR{#1}}}

\newcommand{\EX}[1]{\ensuremath{\mathcal{E}_X\BR{#1}}}
\newcommand{\EXY}[1]{\ensuremath{\mathcal{E}^X_Y\BR{#1}}}
\newcommand{\EXR}[1]{\ensuremath{\mathcal{E}^X\BR{#1}}}

\newcommand{\MK}[1]{\mathsf{mk#1}}

\newcommand{\aside}[1]{\ignorespaces}

\keywords{metaprogramming, code generation, two-level languages,
  let-insertion, mutual recursion}

\begin{document}

\title{\textsf{let} (\textsf{rec}) insertion without Effects, Lights or Magic}
\author{Oleg Kiselyov}
\address{Tohoku University, Japan}
\email{oleg@okmij.org}
\urladdr{http://okmij.org/ftp/}
\author{Jeremy Yallop}
\address{University of Cambridge, UK}
\email{jeremy.yallop@cl.cam.ac.uk}

\begin{abstract}
\emph{Let insertion} in program generation is producing code with
definitions (let-statements). Although definitions precede uses in
generated code, during code generation `uses' come first: we might not
even know a definition is needed until we encounter a reoccurring
expression.  Definitions are thus generated `in hindsight', which
explains why this process is difficult to understand and implement~--
even more so for parameterized, recursive and mutually recursive
definitions.

We have earlier presented an interface for \textit{let(rec)
  insertion}~-- i.e.~for generating (mutually recursive) definitions.
We demonstrated its expressiveness and applications, but not its
implementation, which relied on effects and compiler magic.

We now show how one can understand let insertion, and hence implement
it in plain OCaml.  We give the first denotational semantics of
let(rec)-insertion, which does not rely on any effects at all.  The
formalization has guided the implementation of let(rec) insertion in the
current version of MetaOCaml.
\end{abstract}
\maketitle

\section{Introduction}
\label{s:intro}

Code generation, whether using quasiquotes or code combinators, is
compositional: nested function calls in the generating program lead to
nested expressions in the generated code, and code for larger
expressions is built by incorporating code for sub-expressions
unchanged. Here is an example, with code combinators (used later in the
paper). Suppose |t1| denotes the code ``\texttt{1+2}''; |+
are the combinators to generate addition (resp.,
let- and lambda-expressions). Then
\begin{code}
fnCx. (clet t1 (fun y -> x +% y))
\end{code}
generates the code
\begin{code}
fun x7 -> (let y8=1+2 in (x7 + y8))
\end{code}
(The parentheses show the corresponding nested scopes.)

There is, however, often a need for a sub-expression to generate a
let-statement that should scope over a larger (parent)
expression~\cite{design-and-implementation-ber}~-- e.g.~to avoid
recomputations. Continuing the example, we would like to replace
|clet| above with a combinator |glet| such that
\begin{code}
fnCx. (glet t1 (fun y -> x +% y))
\end{code}
would now generate a more optimal
\begin{code}
let y8=1+2 in fun x7 -> (x7 + y8)
\end{code}
As this example shows, let-insertion is non-compositional: it
scrambles the nesting of generated binding forms, opening the
possibility of generating code with unbound or mistakenly bound
variables~\cite{hygienic-combinators}. It is not a mere possibility:
generating code with unbound variables does occur in practice, and is
difficult to debug, as reported in~\cite{RandIR}.  The
non-compositionality becomes glaring when generating recursive (and
especially mutually recursive) definitions~\cite{genletrec}.

Our aim is to understand the meaning of the let-inserting code
generators, such as |glet| and its more general 
forms |genlet| or |genletrec|.  We have two goals:
designing a type system to statically prevent the generation of ill-scoped
code; and reasoning about programs that generate let(rec)
statements (not just about the code that they generate).

We report work-in-progress towards these goals: a denotational
semantics that for the first time describes what |genlet| and
|genletrec| mean by \emph{themselves}, and in a compositional way.
That is, the apparent intrinsic non-compositionality of let-insertion
described above turns out to be a mere appearance. 

The key idea is virtual let-bindings: whereas
|clet| introduces an ordinary let-binding whose location is fixed,
|glet| generates the code for a fresh variable accompanied by a
virtual binding of that variable. Virtual bindings do not
have (yet) a fixed location: they are attached to the expression that
uses their bound variables and `float up' when their attached
expression is incorporated into a bigger expression. Eventually, when
they reach a point that a metaprogrammer has marked with a dedicated
primitive, virtual bindings are converted to real let-bindings.

Our denotational semantics is executable: it serves as a small
standalone meta\hyp programming system that implements the
previously-proposed interface \cite{genletrec}
for generating mutually recursive
definitions; it is sufficiently complete to express the example
programs used to introduce that interface.
Compositionality let us build the system in pure OCaml using no
effects.
Furthermore, our semantics has already led to improvements and
simplifications to the earlier interface.
\begin{rem}
When saying that our system performs let-insertion without any
effects, we need to clarify what an effect is. After all, even
lambda-abstractions \cite{having-effect} or variable substitution
\cite[\S2.1]{NBEA} may be regarded as effects. Previously,
let-insertion required control effects (realized as delimited control
or continuation-passing transformation)
\cite{Bondorf:1992:IBT:141471.141483,Lawall:1994:CPE:182409.182483}
or, at the very least, state (realized as mutable state or state-passing)
\cite{sumii-hybrid}. The present paper demonstrates that none of these
are needed. As a consequence, if in a generator expression such as
|e1 +
summand generators  |e1| and |e2| may be evaluated
independently or even concurrently, they can be so evaluated even if
either or both perform let-insertion. That let-insertion
does not have to impose an order on evaluation is new and surprising.
\end{rem}

The next section introduces denotational semantics for code
generation, which
\S\ref{s:genlet} extends to support the ordinary let insertion,
and then mutually recursive-let--insertion. \S\ref{s:metaocaml}
describes the realization in the current version of MetaOCaml.
Related work is briefly described in \S\ref{s:related}.

The complete code of the executable semantics along with many examples 
is available online.\footnote{\url{http://okmij.org/ftp/meta-programming/genletrec}} MetaOCaml
(version N111) incorporating let(rec) insertion is available from
Opam.\footnote{\url{https://opam.ocaml.org/}}


\section{Semantics of Code Generation}
\label{s:denot}

As the |Base| calculus
we take the standard call-by-value simply-typed lambda calculus
with constants, ordinary let-expressions and (potentially mutually)
recursive letrec-expressions: think of the most basic,
side-effect--free subset of OCaml.  Figure~\ref{f:base} presents its
syntax.  There, |c0|, |c1|, |c2| and |c3| stand for constants of the
corresponding arity. \aside{Applications of constants to fewer arguments than
their arity are not considered expressions.} The calculus includes
integer and boolean literals (as zero-arity constants), the successor
operation |succ| as an arity-1 constant, and  arithmetic and comparison
operations on integers, of obvious types, as arity-2 constants:
see Fig.~\ref{f:constants}, left column.
The type system with
judgements |'G vee e:t| is entirely standard and elided for brevity.
\aside{in the full version, add the type system figure. Then state the
  theorem that the type system is sound because eval clauses are all
  total. Draw the attention why E[x] rho = rho(x) is total:
  because, in full, E[Gamma |- x:t] rho, rho is in T[Gamma], hence
  x, which must be mentioned in Gamma, is in the domain of rho.
  This is trivial for now: but when we add let-insertion we see how
  this assumption, x is always in the domain of rho, is violated.
  Perhaps state it as a preposition: x is in the domain of rho.
  Incidentally, try to give type to mkvar. It becomes clear how to
  design a type system to prevent scope extrusion, from the type of
  mkvar that makes it total.
  Perhaps the judgement should be instead of Gamma |- e:t,
  $e^\Gamma: t$. See Pientka's et al. paper in POPL 2022.
}

\begin{figure}[t]
\begin{tabular}[Ct]{ll}
Variables & |x,y,z,u,f,n,r|\ldots \\
Types &
\begin{code}
t ::= int | bool | t -> t
\end{code}
\\
Expressions &
\begin{code}
e ::= x | c0 | c1 e | c2 e e | c3 e e e 
   | fnx. e | e e 
   | if e then e else e 
   | let x=e in e
   | let rec x=e and x=e ... in e
\end{code}
\\
Values &
\begin{code}
v ::= c0 | fnx. e
\end{code}
\\
\end{tabular}
\caption{Syntax of the base calculus; \textsf{c$_i$} are constants 
of arity $i$: see Fig.~\ref{f:constants}, left column.
}
\label{f:base}
\end{figure}

\begin{figure}
\begin{tabular}[Ct]{l@{\hskip 3em}r}
\begin{code}
0,1,2,3,...  : int
true, false: bool
succ: int -> int
+: int -> int -> int
=: int -> int -> bool
\end{code}
&
\begin{code}
cint: int -> int code
cbool: bool -> bool code
+%: int code -> int code -> int code
=%: int code -> int code -> bool code
cif: bool code -> $t$ code -> $t$ code -> $t$ code
fnC: ($t_1$ code -> $t_2$ code) -> ($t_1\to t_2$) code
@: ($t_1\to t_2$) code -> $t_1$ code -> $t_2$ code
clet: $t_1$ code -> ($t_1$ code -> $t_2$ code) -> $t_2$ code
\end{code}
\end{tabular}
\caption{Constants of Base (left column) and its extension Codec
  (right column) with their types and arities: the arity of a constant
  is the number of arrows in its type.  Although the constants may
  have function types, they are not expressions, unless used with the
  right number of arguments.  The metavariable $t$ stands for any
  type. For arity-2 constants such as  \textsf{+} and \textsf{=}, we
  use infix notation. We write expressions with the constant
  $\underline\lambda$, viz.,
  $\underline\lambda\mathsf{(\lam{x}e)}$, as
  $\mathsf{\quant{\underline\lambda}{x}e}$.
  We silently add other arithmetic and comparison constants and
  code combinators, similar to \textsf{+} and \textsf{=}.  }
\label{f:constants}
\end{figure}

A different way of presenting the calculus and its type system is in
the form of an OCaml signature (Appendix~\ref{s:sigs}), which helps make
the semantics executable.

Here are some sample expressions:
\begin{code}
t1   := 1 + 2
sq   := fnx. x * x
gib5 := fnx.fny.
  let rec loop n = 
      if n=0 then x else if n=1 then y else 
      loop (n-1) + loop (n-2)
  in loop 5
\end{code}
The notation |name := exp| is not part of the calculus; it is used to
attach a name to an expression for easy reference. The function
|gib5| computes the 5th element of the Fibonacci sequence whose first
two elements are given as arguments.

The |Base| calculus both represents the code that we generate and
serves as the core of the generating code. For generation, we extend
|Base| with an additional family of types |t code| whose values
represent generated |Base| expressions of type |t|. We also
add the means of producing these code values: constants (a.k.a. code-generating
combinators) in the right column of Fig.\ref{f:constants}.
Below are some expressions in this
extension of |Base|, called |Codec|; each expression serves as a generator of
the corresponding earlier |Base| expression:
\begin{code}
ct1   := cint 1 +% cint 2
csq   := fnCx. x *% x
cgib5 := fnCx.fnCy. 
   let rec loop n =
       if n=0 then x else if n=1 then y else
       loop (n-1) +% loop (n-2)
   in loop 5
\end{code}
Here |cint| generates the code of an
integer literal; |+
combines the code of summands to produce the code of an addition expression;
|fnCx.body| generates the code of a function given a generator
for its body, the variable |x| within the expression |body|
representing the bound variable in the (to be) generated function.

In what sense |csq| and |cgib5| respectively represent |sq| and |gib5| will
be clear after we describe the semantics of the calculus. 

\subsection{Semantics of \textsf{Base}}

We consider two denotational semantics of |Base|, to be indexed by the
subscripts |R| (for `run') or |S| (for `show'); 
|X| (or omitted subscript) stands for either. 

First, notated by the subscript |R|, is the
standard Scott-Strachey semantics for a typed Church-style calculus, 
with one small wrinkle. Its semantic domains and the interpretation
\TR{-} of its types
are standard:
\[
\TR{int} = \mathbb{Z}_\bot
\qquad
\TR{bool} = \{\mathtt{tt},\mathtt{ff},\bot\}
\qquad
\TR{t_1\to t_2} = \TR{t_1} \to \TR{t_2}
\]
If $A$ is a set and $B$ is a domain, $A\to B$ is a continuous map from
$A$ to $B$, which is also a domain. We also introduce
$\Nom$ as a countably infinite set of names and $\Loc$ as a set of finite
sequences of small (i.e.~bounded) numbers, for which we adopt the OCaml list
notation. Such a sequence can also be considered a name. Therefore,
we take $\Loc\subset\Nom$ and treat $\Loc$ as names, distinct from the names
appearing in source |Base| terms.

The semantic function $\mathcal{E}_X\big[\Gamma\vdash
  \mathsf{e}:\mathsf{t}\big] \in \mathcal{D}_X[\mathsf{t}]$
gives meaning to (the type
derivation of) a potentially open expression |e|, where 
$\mathcal{D}_X[\mathsf{t}]$ along with auxiliary domains is defined as follows.
(When writing semantic functions, we shall show only the expression rather than
its entire type derivation, and often
elide $\Gamma$ and the type annotations to avoid clutter.)
\aside{Actually, I should have annotated D with Gamma, and wrote
$D_X^\Gamma[t] = T[\Gamma]\to \Loc \to T[t]$ where $T[\Gamma]$ is a
  record, which can be regarded as a finite map, but the names $\Nom$
  should be in dom(Gamma)
  But here we are sloppy and introduce partiality.
}
\begin{tabular}[C]{Ml}
\mathcal{D}_X[\mathsf{t}] = {Env}_X \to
\Loc\to \TX{t}
\\
{Env}_X = \Nom\rightarrowtail \TX{t}
\end{tabular}
where $A\rightarrowtail B$ is a finite map from $A$ to $B$. If $\rho$
is such a map, $\rho[k{\to} v]$ is its extension, $\rho\mid_{\not=k}$
is its restriction (removing the association for $k$) and
$\mathrm{dom}\,\rho$ is its domain; $\varnothing$ is the empty map.
We write $\mathsf{modify}\,\rho\,k\,u$ the modification of the element
at key $k$ by an update function $u$, that is: $\rho[k{\to}u\,\rho(k)]$.

The semantic function 
$\mathcal{M}_X\BR{e:t} \in \TX{t}$ gives meaning to
programs (i.e.~to type derivations of closed expressions):
\[
\mathcal{M}_X\BR{e} = \mathcal{E}_X\BR{e}\ \varnothing\  \mathsf{[]} 
\]

The semantic rules are almost entirely standard:
the extra $\Loc$ argument, written as |'l|, is the wrinkle.
Here is the rule for abstraction:
\begin{code}
ER[fnx.e]! 'r 'l = $\lam{x}\mathcal{E}_R\BR{e}$ 'r[x->$x$] (1::'l)
\end{code}
For now, |'l| is not actually used, and might as well be
absent. It will help soon, in \S\ref{s:sem-codec}. 
It will also help to re-write the semantic
rules in the form shown in Fig.~\ref{f:base-semantics}, with the `composition'
and variable reference rules common to all semantics, and the
semantic-specific strict |mk| functions.
For instance, $\MK{lam}_R$ takes the variable name and the
denotation of a (generally open) expression and constructs the
denotation of a lambda-abstraction. The re-written rules make it very
clear that the denotation of, say, an abstraction is constructed from the
denotation of the abstraction body and the name of the abstracted
variable. \aside{needs the rules for constants like e1+e2, and for if
and let rec}

\begin{figure}
\begin{array}[C]{MrMl}
\tabpause{(a) composition rules}
\EX{i} &= \MK{int}_X\ \mathsf{i}
\\
\EX{e_1 + e_2} &= \MK{add}_X\ \EX{e_1}\ \EX{e_2}
\\
\EX{x} &= \MK{var}\ \mathsf{x}
\\
\EX{\lam{x}e} &= \MK{lam}_X\ \mathsf{x}\ \EX{e}
\\
\EX{e_1\ e_2} &= \MK{app}_X\ \EX{e_1}\ \EX{e_2}
\\
\EX{let\ x=e_1\ in\ e_2} &= \MK{let}_X\ \mathsf{x}\ \EX{e_1}\ \EX{e_2}
\\
\tabpause{(b) semantics of variable reference (generic)}
\MK{var}\ \mathsf{x} &= \lam{\rho\ell}\rho(\mathsf{x})
\\
\tabpause{(c) $\mathsf{mk}$ functions for the |R| semantics}
\MK{int}_R\ i &= \lam{\rho\ell}i
\\
\MK{add}_R\ d_1\ d_2 &= 
 \lam{\rho\ell} (d_1\,\rho\, (1::\ell)) + (d_2\,\rho\,(2::\ell))
\\
\MK{lam}_R\ \mathsf{v}\ d &= 
\lam{\rho\ell}\lam{x} d\,\rho[\mathsf{v}{\to} x]\, (1::\ell)
\\
\MK{app}_R\ d_1\ d_2 &= 
 \lam{\rho\ell} (d_1\,\rho\, (1::\ell))\ (d_2\,\rho\,(2::\ell))
\\
\MK{let}_R\ \mathsf{v}\ d_1\ d_2 &= 
 \lam{\rho\ell}(\lam{x}d_2\,\rho[\mathsf{v}{\to}x]\,(2::\ell))\ 
               (d_1\,\rho\,(1::\ell))
\\
\tabpause{(d) $\mathsf{mk}$ functions for the |S| semantics}
\MK{int}_S\ i &= \lam{\rho\ell}\textsf{printf"\%d"}\ i
\\
\MK{add}_S\ d_1\ d_2 &= 
 \lam{\rho\ell}\textsf{printf"(\%s+\%s)"}\
 (d_1\,\rho\, (1::\ell))\ (d_2\,\rho\,(2::\ell))
\\
\MK{lam}_S\ \mathsf{v}\ d &= 
 \lam{\rho\ell}\textsf{printf"($\lambda$\%s. \%s)"}\ \mathsf{v}\
 (d\,\rho[\mathsf{v}{\to} \mathsf{v}]\, (1::\ell))
\\
\MK{app}_S\ d_1\ d_2 &= 
 \lam{\rho\ell}\textsf{printf"(\%s \%s)"}\
 (d_1\,\rho\, (1::\ell))\ (d_2\,\rho\,(2::\ell))
\\
\MK{let}_S\ \mathsf{v}\ d_1\ d_2 &= 
 \lam{\rho\ell}\textsf{printf"(let \%s=\%s in \%s)"}\ \mathsf{v}\\
 & \hfill
               (d_1\,\rho\,(1::\ell))\
               (d_2\,\rho[\mathsf{v}{\to}\mathsf{v}]\,(2::\ell))
\end{array}
\caption{Semantics of Base}
\label{f:base-semantics}
\end{figure}

The semantics of |Base| just given could rightly be called
`extensional'. We also have an `intensional' |Base| semantics, 
notated by the superscript |S|, which maps
an expression to its symbolic form (a string, for example). Here 
$\TS{t}$ is always a string and the functions 
$\MK{lam}_S$ etc.~build strings (see Fig.\ref{f:base-semantics}(d)).
|S| is a
trivially compositional, \textit{bona fide} denotational semantics, and even
mentioned as such by \cite{Mosses-denot}. Usually it is quite
useless~-- but not here.

\subsection{Semantics of \textsf{Codec}}
\label{s:sem-codec}

The semantics $\mathcal{M}^X_Y[-]$ and $\EXY{-}$ of |Codec|
is an extension of a semantics of |Base|. There are now two indices: the
subscript |Y| notates the semantics (|R| or |S|) of the generator,
whereas the superscript |X| labels the semantics used for the
generated code. (For a two-level language, |Y| being |R| is the most
useful variant; we often leave this implicit and drop |Y|.) 
The semantic domain of |t code| is (for now; it will be extended when
we come to let- and let-rec-insertion):
\[
\TUX{t\ code} = \mathcal{D}_X[\mathsf{t}]
\]
A |t code| value represents a potentially open (think of generating
function bodies) |Base| expression |e| of type |t|. Its meaning 
is, therefore, $\mathcal{E}_X[\mathsf{e}]$: the meaning the |Base|
semantics |X| gives to it.
\aside{If we annotated DX with Gamma, as we should, then the question
  arises: which Gamma to use in the above equation? The answer is to
  annotate t code with Gamma: contextual modal type theory. But we can
  omit then, and prove that one can do things safely at the expense of
  a dynamic check (at code generation time!) and emitting good error messages.
}

Since |Codec| is an extension of |Base|, its semantics $\EXY{-}$ is an
extension of $\mathcal{E}_Y\BR{-}$ with the rules for 
the meaning of constant expressions of type |t code|, such as the
following rules for generation of an integer literal, abstraction, application
and let-expression. Generating addition, comparison, if-expression
is similar. As far as variable references are concerned,
$\EXY{x} = \mathcal{E}_Y\BR{x} = \MK{var}\ \mathsf{x}$.
\begin{align*}
\EXY{\underline{int}\ i}\rho\ell &= \MK{int}_X\ \mathsf{i}
\\
\EXY{\quant{\underline\lambda}{x}e}\rho\ell &= 
 \MK{lam}_X\ \ell\
 (\EXY{e}\, \rho[\mathsf{x}\to \MK{var}\ \ell]\ (1::\ell))
\\
\EXY{e_1\ \underline{@}\ e_2}\rho\ell &= 
\MK{app}_X\ (\EXY{e_1}\,\rho\,(1::l))\ (\EXY{e_2}\,\rho\,(2::l))
\\
\EXY{clet\ e_1\ e_2}\rho\ell &=
\MK{let}_X\ \ell\ (\EXY{e_1}\,\rho\,(1::l))\
((\EXY{e_2}\,\rho\,(2::l))\ (\MK{var}\ \ell))
\end{align*}
When we generate an abstraction or a let-expression, the current |'l| 
acts as the fresh name for the (to be) bound variable. Recall that the
function $\MK{lam}$ (Fig.\ref{f:base-semantics}) takes a variable name
and the denotation for the abstraction body and gives the denotation
for the abstraction. Thus the role of |'l|, the only non-standard
aspect of the |Base| semantics, is to serve as a deterministic name
generator: the fresh name to be used in an expression is the path from
the root of the B{\"o}hm tree.

If we use the |R| semantics for the generated code (that is, choose
|X| to be |R|) we see that 
|MRR[ct1]| is exactly |MR[t1]!| (which is
the integer 3), |MRR[csq]!| and |MR[sq]!| both mean the
squaring function, and |MRR[cgib5]!| and |MR[gib5]!| both
mean the function that takes two arguments |x| and |y| and returns the sum of
5 copies of |y| and 3 copies of |x|.

If we use the |S| semantics, $\mathcal{M}^S_R[\mathsf{ct1}]$ and
$\mathcal{M}_S[\mathsf{t1}]$ still
coincide (both mean the string |1+2|). $\mathcal{M}^S_R[\mathsf{csq}]$
and $\mathcal{M}_S[\mathsf{sq}]$ are generally different but 
$\alpha$-equivalent 
lambda\hyp expression strings. Whereas 
$\mathcal{M}_S[\mathsf{gib5}]$ is the string of the
|gib5| code (potentially $\alpha$-converted), 
$\mathcal{M}^S_R[\mathsf{cgib5}]$ is the string
\begin{code}
fnx.fny. (((y + x) + y) + (y + x)) + ((y + x) + y)
\end{code}
It is an `optimized' version of |gib5|, in the sense that the loop is
unrolled; however, it contains several instances of code
duplication. Avoiding this code duplication is where let-insertion
comes in.

\section{Let-insertion}
\label{s:genlet}

To support let-insertion, we add to |Codec| two more forms:
|let locus l in e| and |genlet l e_m e|. The former, like
the ordinary let, binds the so-called locus variable |l| in |e|.
In the expression |genlet l e_m e|, |l| is a locus variable (previously bound
by |let locus|), |e_m| is a so-called memo key (for now, an |int|
expression) and |e| is a |t code| expression (|t| is a |Base| type).
Roughly, |e| generates the right-hand-side of the let-binding, |l|
tells where to insert it, and the memo key instructs which |e| are to
be shared. We describe the |genlet| arguments in more detail after the
example, the \emph{slightly} adjusted |cgib5|:
\begin{code}
clgib5 := fnCx.fnCy. let locus l in
   let rec loop n =
     if n=0 then x else if n=1 then y else
     genlet l (n-1) (loop (n-1)) +% genlet l (n-2) (loop (n-2))
   in loop 5
\end{code}
To a first approximation, one may think of |genlet l e_m e| as
generating |let z=c in z| where |z| is fresh and |c| is the code produced by
the expression |e|. Such `let-expansion' is useless, however. It becomes
more useful when the binding |let z=c in| is actually placed somewhere
`higher' in the overall generated code. The form |let locus l| marks
that `higher' place where the bindings produced by |genlet| are to be
placed. Since let-insertion is very common, different parts of the
generator may do their own let-insertions at different places; the
locus variable |l| is to connect |genlet| with its corresponding 
\textsf{let locus}.
Thus intuitively, |genlet l e_m e| will insert the 
|let z=c in| at the place marked by \textsf{let locus l} and return the code
of the bound variable |z| (which is distinct from any other
variables in the code). 

Placing let-bindings `higher'
in the code is useful because they may be shared. The memo key defines
the equivalence classes: expressions with the same
memo key are to be shared. Therefore, if |genlet l e_m e| finds that there
is already a let-binding produced by an earlier |genlet| with
the same |l| and the memo key, |genlet l e_m e| returns the code of the earlier
bound variable.

Using the semantics of these operations, explained below, we can
see that whereas
|MRR[clgib5]!| remains the same as |MR[gib5]!|,
|MSR[clgib5]!| is the string
\begin{code}
fnx.fny. let z = y in let u = x in 
         let v = z + u in let w = v + z in 
         let x6 = w + v in x6 + w
\end{code}
which is indeed an optimized version of |gib5|, without either loops
or duplication.

\subsection{Semantics of let-insertion}

The key point was that the binding introduced by |genlet l e_m e| is
not yet placed; its placement will be decided only later, upon
seeing the corresponding |let locus|. For now, the |genlet|'s binding
is `floating'~-- we say `virtual'. To accommodate virtual bindings
we extend the semantics domain of |t code|: it is now a tuple, whose
first component is the earlier semantic domain of code values (\S\ref{s:sem-codec}),
and whose second component is the virtual bindings. Formally, the semantic domain is:
\begin{tabular}[C]{MrMl}
\TUX{t\ code} =& 
  \mathcal{D}_X[\mathsf{t}] \times (\Loc\rightarrowtail\mathcal{V}^X)
\\[0.7ex]
\mathcal{V}^X =& (\mathcal{K}\times \mathcal{K}) {Set}\times (\mathcal{K}\rightarrowtail \mathcal{B}^X)
\\[0.7ex]
\mathcal{B}^X =& \Nom \times \mathcal{D}_X[\mathsf{t}]
      \times \Nom Set
\end{tabular}

Virtual bindings are indexed by the locus where they will be actually
inserted. Furthermore, virtual bindings with the same locus $l$ and the
same memo key $k$ belong to the same equivalence class. We take
the locus to be an element of $\Loc$ and introduce the set
$\mathcal{K}$ of memo keys. All in all,
virtual bindings is a finite map $\Loc\rightarrowtail\mathcal{V}^X$,
where $\mathcal{V}^X$ describes virtual bindings with the same locus.
If $\nu$ is such a map, we take $\nu(l)=\varnothing$ if $l\not\in
\mathrm{dom}(\nu)$.
Further thought, considering 
|genlet l 2 (cint 3 +
virtual bindings have to be (partially) ordered. (We shall see an
example soon.) Thus the elements of
$\mathcal{V}^X$ are tuples $\langle R,b \rangle$ where $R$ is a preorder on
$\mathcal{K}$ and $b$ is a finite map
$\mathcal{K}\rightarrowtail \mathcal{B}^X$. Here, $\mathcal{B}^X$
describes one equivalence class of virtual bindings: a tuple
$\langle n,d,\overline n\rangle$ where $n$ is the name to bind,
$d$ is the ($\EX{-}$ denotation of the) expression to which $n$
will be bound to, and
$\overline n$ is the set of names: names equivalent to $n$.

The earlier semantic rules for $\EXY{-}$ dealing with code generation
have to be amended to account for the extended semantic domain: 
$\EXY{\underline{int}\ i}$ produces the empty virtual binding,
and the other rules propagate the virtual bindings of their
subexpressions, merging as needed:
\begin{align*}
\EXY{\underline{int}\ i}\rho\ell &= 
\langle \MK{int}_X\ \mathsf{i},\ \varnothing\rangle
\\
\EXY{\quant{\underline\lambda}{x}e}\rho\ell &= 
 \langle \MK{lam}_X\ \ell\ d,\ \nu\rangle\qquad \textrm{where}\\
& \langle d,\nu \rangle = 
 \EXY{e}\, \rho[\mathsf{x}\to \langle \MK{var}\ \ell,\varnothing\rangle]\ 
           (1::\ell)
\\
\EXY{e_1\ \underline{@}\ e_2}\rho\ell &= 
\langle \MK{app}_X\ d_1\ d_2,\ \mathsf{merge}\ \nu_1\ \nu_2\rangle
\qquad \textrm{where}\\
& \langle d_1,\nu_1 \rangle = \EXY{e_1}\,\rho\,(1::l)\\
& \langle d_2,\nu_2 \rangle = \EXY{e_2}\,\rho\,(2::l)
\end{align*}
The operation $\mathsf{merge}$ for virtual
bindings is described later.

As we said earlier, |genlet l e_m e| generates a fresh name to which
the code produced by |e| will eventually be bound; that fresh name is
accompanied by the new virtual binding of that name to the result of
|e|. Here, |e| is a |t code| expression: the generator of the
expression to bind. It itself may be accompanied by virtual
bindings. The new binding added by |genlet l e_m e| may in general
depend upon those bindings, and hence has to be added as `greater' in
the preorder $R$. (In contrast, |e_m| is an |int| rather than 
an |int code| expression, and so its denotation $\EXY{e_m}\,\rho\,(1::\ell)$,
written as $k$ below, is (when |Y| is |R|) just an integer: in
general, an element of $\mathcal{K}$.)
\begin{align*}
\EXY{genlet\ l\ e_m\ e}\rho\ell &= \langle\:\MK{var}\,\ell,\ 
\mathsf{modify}\,\nu\,\rho(\mathsf{l})\,
(\mathsf{addb}\,k\, \langle\ell,d_b\rangle)\:\rangle\qquad \textrm{where}\\
 k & = \EXY{e_m}\,\rho\,(1::\ell)\\
 \langle d_b,\nu\rangle & = \EXY{e}\,\rho\,(2::\ell)
\end{align*}
(If |e_m| or |e| diverge, so does |genlet l e_m e|.)

The semantic function $\mathsf{addb}\, k\, \langle n,d\rangle\, v$
adds a new virtual binding of $n$ with $d$ to the equivalence class
$k$ of the virtual bindings $v\in \mathcal{V}^X$ with the same locus.
Recall that the virtual bindings $v$ is a pair, of preorder $R$ and
the set of equivalence classes $b$, indexed by the memo key. There are
two cases to consider: if $b$ already includes the equivalence class
for $k$, we add $n$ to that class (and disregard the right-hand-side
$d$ since we already have an equivalent one). Otherwise, we add to $b$
the new equivalence class for $k$ containing just the binding of $n$
to $d$, and update the preorder $R$ so that $k$ becomes the `latest'.
\begin{array}[C]{l}
\mathsf{addb}: \mathcal{K}\to (\Nom\times \mathcal{D}_X[t])\to
\mathcal{V}^X\to\mathcal{V}^X
\\
\mathsf{addb}\ k\ \langle n,d\rangle\ \langle R,b\rangle =
\begin{cases}
\langle R, b[k{\to}\langle n',d',\overline{n'}\cup \{n\}\rangle]\rangle \\
 \qquad \textrm{if }
b(k) = \langle n',d',\overline{n'}\rangle
\\
\langle R \cup \{(k,k)\}
\cup \{\langle k',k\rangle \mid k'\in \mathrm{dom}\, R\},\
 b[k{\to}\langle n,d,\varnothing\rangle ]\rangle \\
 \qquad \textrm{if }
k\not\in \mathrm{dom}\, b 
\end{cases}
\end{array}

The form |let locus l in e| converts the virtual bindings
for the locus |l|
produced by |e| into real let-bindings. To a
first approximation, the conversion can be understood as turning the
sequence of virtual bindings into nested let-expressions. We should
mind the dependency among the bindings and nest the 
let-expressions in the `right' order.
\begin{align*}
\EXY{let\ locus\ l\ in\ e}\rho\ell &= 
\langle \mathsf{bind} (\mathsf{ordered}\,\nu(\ell))\,d,\ 
\nu\mid_{\not=\ell}\rangle
\qquad \textrm{where}\\
&\langle d,\nu\rangle = \EXY{e}\,\rho[\mathsf{l}\to\ell]\,(1::\ell)
\end{align*}
The semantic function $\mathsf{ordered}: \mathcal{V}^X\to
\mathcal{B}^X Seq$ converts $\langle R,b\rangle$ to a sequence of bindings
$\mathcal{B}^X$ in an order consistent with $R$. The semantic
function |bind| converts a sequence of bindings to nested
let-expressions.
 
\begin{tabular}[C]{Ml}
\mathsf{bind}
       [\langle n_1,d_1,\overline{n_1}\rangle,\langle n_2,d_2,\overline{n_2}\rangle,\ldots]\, d
       =\\
\quad
 \mathsf{mklet}_X\, n_1\ d_1\ (\mathsf{subst}\, n_1\,
 \overline{n_1}
 (\mathsf{mklet}_X\, n_2\ d_2\ (\mathsf{subst}\, n_2\, \overline{n_2}
 (\ldots d))))
\end{tabular}

\noindent
Recall, the semantic function |mklet| (see Fig.\ref{f:base-semantics}) 
builds the denotation of
|let x=e in e'| from the variable name |x|, 
$\mathcal{E}_X[\mathsf e]$ and
$\mathcal{E}_X[\mathsf{e'}]$. In a virtual binding $\langle n,d,\overline n\rangle$, the set 
$\overline n$ contains the variables other than $n$ in the same
equivalence class with it. They are all substituted with $n$:
\begin{tabular}[C]{Ml}
\mathsf{subst}\, n\,\overline{n}\ d = 
\lam{\rho\ell} d\ \rho[(n'{\to} n \mid n'\in \overline n)]\ \ell
\end{tabular}

The semantics function $\mathsf{merge}\, \nu_1\,\nu_2$ mentioned earlier
merges the virtual bindings $\nu_1$ and $\nu_2$,
by using $\mathsf{addb}$ to add one-by-one the virtual bindings of 
$\nu_2$ to $\nu_1$, in an order consistent with the $R$ preorder of 
$\nu_2$.

An example featuring code duplication and nesting should show how
everything fits together. As a warm-up, a generator without |genlet|:
\begin{code}
let locus l in 
 let x = (cint 6 +% cint 7) in
 ((x +% cint 20) *% (x +% cint 30)) /% cint 100
\end{code}
has the meaning in the |MSR[-]!| semantics as a string
\begin{code}
(((6 + 7) + 20) * ((6 + 7) + 30)) / 100
\end{code}
with evidently duplicate code. We can eliminate the duplication by putting
let-expressions into the generated code, using |genlet|:
\begin{code}[numbers=left]
let locus l in 
 let x = genlet l 1 
             (cint 6 +% cint 7) in
 (genlet l 2 (x +% cint 20) 
   *% 
   genlet l 3 (x +% cint 30)) 
  /% cint 100
\end{code}
The three |genlet| expressions are each in their own equivalence class
and hence pass distinct memo keys as their second arguments.
(Such usage can be automated~-- in fact, it has been, in MetaOCaml.)
The code is typeset so that each notable expression is on its own line
for easy reference. Furthermore, we write $\ell_i$ for the B{\"o}hm
tree location (an element of $\Loc$) corresponding to line $i$.

The denotation is computed as follows. Let $\rho_0$ be the initial environment
and $\rho=\rho_0[\mathsf{l}\to \ell_1]$.
We also assume $\MK{mul}$ and
$\MK{div}$ functions similar to $\MK{add}$.
\begin{array}[C]{l}
\EXR{(\underline{int}\,6\ \underline{+}\ \underline{int}\,7)}\rho\ell_3 =  
\langle d_3,\ \varnothing\rangle
\qquad\textrm{where }
d_3 = \MK{add}_X\,(\MK{int}_X\,6)\,(\MK{int}_X\,7)
\hskip 2em
\\
\EXR{genlet\,l\,1\,
 (\underline{int}\,6\ \underline{+}\ \underline{int}\,7)}\rho\ell_2 =
\langle \MK{var}_X\,\ell_2,\ \{\ell_1\to v_2\}\rangle
\quad\textrm{where}
\\
\hfill
v_2 = \mathsf{addb}\,1 \langle \ell_2,d_3\rangle\,\varnothing =
\langle \{(1,1)\},
           \{1\to \langle \ell_2,d_3,\varnothing\rangle\}\rangle
\\
\tabpause{This was the denotation of the expression
the variable \textsf{x} is bound to. The variable is then used twice,
on lines 4 and 6. In the following we take $\rho_1$ to be $\rho$
extended with the binding for \textsf{x}.}
\EXR{genlet\,l\,2\,(x\ \underline{+}\ \underline{int}\,20)}\rho_1\ell_4 =
\langle \MK{var}_X\,\ell_4,\ \{\ell_1\to v_4\}\rangle
\quad\textrm{where}
\\
\hfill
d_4 = \MK{add}_X\ (\MK{var}_X\,\ell_2)\ (\MK{int}_X\,20)\\
\hfill
v_4 = \mathsf{addb}\,2\,\langle \ell_4,d_4\rangle\,v_2 =
\langle \{(2,2),(1,1),(1,2)\},
           \{1\to \langle \ell_2,d_3,\varnothing\rangle,
           2\to \langle \ell_4,d_4,\varnothing\rangle\}\rangle
\\
\EXR{genlet\,l\,3\,(x\ \underline{+}\ \underline{int}\,30)}\rho_1\ell_6 =
\langle \MK{var}_X\,\ell_6,\ \{\ell_1\to v_6\}\rangle
\quad\textrm{where}
\\
\hfill
d_6 = \MK{add}_X\ (\MK{var}_X\,\ell_2)\ (\MK{int}_X\,30)\\
\hfill
v_6 = \mathsf{addb}\,3\,\langle \ell_6,d_6\rangle\,v_2 =
\langle \{(3,3),(1,1),(1,3)\},
           \{1\to \langle \ell_2,d_3,\varnothing\rangle,
           3\to \langle \ell_6,d_6,\varnothing\rangle\}\rangle
\\
\tabpause{When computing the denotation for the product expression,
we have to merge the virtual bindings $v_4$ and $v_6$}
\EXR{\ldots\ \underline{*}\ \ldots}\rho_1\ell_5 =
\langle \MK{mul}_X\,(\MK{var}_X\,\ell_4)\,(\MK{var}_X\,\ell_6),\ 
\{\ell_1\to v_5\}\rangle
\quad\textrm{where}
\\
\qquad
v_5 = 
\langle \{(3,3),(2,2),(1,1),(1,3),(1,2),(2,3)\},\\
\hfill
           \{1\to \langle \ell_2,d_3,\varnothing\rangle,
           2\to \langle \ell_4,d_4,\varnothing\rangle,
           3\to \langle \ell_6,d_6,\varnothing\rangle\}\rangle
\\
\tabpause{The merged virtual bindings propagate to the denotation of
  the division expression, and converted to real let-bindings by
  let-locus. The whole expression thus has the denotation}
\EXR{let\ locus\ l\ in\ \ldots}\rho_o\ell_1 =\langle\\
\quad
\MK{let}_X\,\ell_2\,(\MK{add}_X\,(\MK{int}_X\,6)\,(\MK{int}_X\,7))\ (
\\ \quad
\MK{let}_X\,\ell_4\,(\MK{add}_X\ (\MK{var}_X\,\ell_2)\ (\MK{int}_X\,20))\ (
\\ \quad
\MK{let}_X\,\ell_6\,(\MK{add}_X\ (\MK{var}_X\,\ell_2)\ (\MK{int}_X\,30))\ (
\\ \qquad
\MK{div}_X\
(\MK{mul}_X\,(\MK{var}_X\,\ell_4)\,(\MK{var}_X\,\ell_6))\
(\MK{int}_X\,100)))),\ \varnothing\rangle
\end{array}

\bigskip
The interface for |genlet| described above differs from our previous
proposal \cite{genletrec} in that here we combine memoization and
let-insertion. Although both memoization and let-insertion are usually
implemented in terms of effects, we have used no effects at all.

If |let locus| in |clgib5| is positioned above |fnCy....|, so called
scope-extrusion occurs, resulting in the generated code 
with have unbound variables (as we can verify in our semantics).
It is the subject of ongoing work to develop a type system to
statically prevent such problems.

\subsection{Generating (mutually) recursive definitions}
\label{s:genletrec}

It turns out that |genletrec| for generating (mutually) recursive
definitions presented by~\cite{genletrec} is a minor variant of the
above |genlet|, with almost the same semantics. The syntax is the
same: |genletrec l e_m e| for requesting the binding (and obtaining
the name of the to-be-bound variable) and |let rec locus l in e| for
actually generating let-rec statements at that locus. 

Here is the example: specializing the Ackermann function challenge by
Neil Jones. The |ack2| below is the two-argument Ackermann function
|ack| partially applied to two:
\begin{code}
ack2 := 
  let rec ack = fnm.fnn.
    if m=0 then n+1 else
    if n=0 then ack (m-1) 1 else
    ack (m-1) (ack m (n-1))
  in ack 2
\end{code}
Below is the generator of |ack2|, matching |ack2| in form.
\begin{code}
cack2 := 
  let rec locus l in 
  let rec ack = fnm.fnC n.
     if m=0 then n+% (cint 1) else
     cif (n =% _cint 0)
       (genletrec l (m-1) (ack (m-1)) @ cint 1)
       (genletrec l (m-1) (ack (m-1))  @  (genletrec l m  (ack m) @ (n-%cint 1)))
   in genletrec l 2 (ack 2)
\end{code}
Here |genletrec| is needed not just to obtain optimal code; without
|genletrec| we get no code at all: divergence. Using the semantics
below, |MSR[cack2]!| gives the following code
\begin{code}
let rec x = fnu. if u = 0 then y 1 else y (x (u - 1))
and y = fnv. if v = 0 then z 1 else z (y (v - 1)) 
and z = fnw. w + 1 in x
\end{code}

Semantically, |genletrec| is also close to |genlet|: both return the 
name of a to-be-bound variable, accompanied by a virtual binding.
Whereas for |genlet l e_m m| the binding associates the name to the
expression produced by |e| (the generated code), for 
|genletrec l e_m e| the virtual binding associates the name to |e| itself
(one may say, unevaluated generator expression). Thus |genletrec| is
even lazier, introducing even more virtual bindings. Formally,
we extend $\mathcal{B}^X$ to
\begin{tabular}[C]{Ml}
\mathcal{B}'^X = 
 \Nom \times (\mathcal{D}_X[\mathsf{t}] + \TUX{t\ code})
      \times \Nom Set
\end{tabular}
where $A+B$ is to be read as a disjoint union (with tags
$\mathtt{inl}$ and $\mathtt{inr}$) with a separately added $\bot$. The
left summand $\mathcal{D}_X[\mathsf{t}]$ is inherited from
$\mathcal{B}^X$: the meaning of the code for the right-hand-side of
the binding. The right summand $\TUX{t\ code}$ is (the approximation of)
the code for the right-hand-side. The $\TUX{t\ code}$ definition
hence becomes recursive and has to be understood as the solution to
the domain equation.  \aside{Typically, genletrec occurs within a
  normal let rec. The right-hand-side may include recursive calls. So,
  the right-hand-side is an approximation, in the information order,
  of that recursive invocation. letrec will find the fixpoint, the
  limit of the chain of approximations. In the full paper, we should
  greatly expand on this.} Then
\begin{align*}
\EXY{genletrec\ l\ e_m\ e}\rho\ell &= 
\langle \MK{var}\,\ell,\ 
\varnothing[l{\to} \mathsf{addb}\,
k\, \langle\ell,\mathtt{inr}\ \EXY{e}\rho\,(2::\ell)\rangle\, 
\varnothing]\rangle
\qquad\textrm{where}\\
&
 l = \rho(\mathsf{l})\\
&
 k = \EXY{e_m}\,\rho\,(1::\ell)
\end{align*}

Virtual bindings now contain yet to be evaluated generator expressions
for the right-hand-side of bindings. To generate the
let-rec-statement, we have to evaluate them~-- which may create more
virtual bindings, which have to be merged and again evaluated (the
evaluation may also diverge). This complex process of evaluation is
called canonicalization, and performed by the semantic function
$\mathsf{canon}\,\nu\,l$ which takes virtual bindings and the locus
$l$ and returns updated bindings $\nu'$ such that $\nu'(l)$ are all
canonical: each $\mathcal{B}'^X$ is actually $\mathcal{B}^X$. The
function is the least fixpoint of the following recursive equation.
\begin{tabular}[C]{Ml}
\mathsf{canon}\,\nu\, l =
\begin{cases}
\nu \textrm{ if $\nu(l)$ are all canonical}
\\
\mathsf{canon}\, (\mathsf{merge}\ \nu[l{\to}\langle R,b'\rangle]\ \nu'')\, l \textrm{ where }\\
\qquad \langle R,b\rangle = \nu(l) 
\\
\qquad
       \langle n,\mathtt{inr}\ d',\overline n\rangle = b(k), \quad
       k \in \mathrm{dom}(b)
\\
\qquad
       \langle d,\nu''\rangle = d'
\\
\qquad
    b' = b[k{\to}\langle n,\mathtt{inl}\ d,\overline n\rangle]
\end{cases}
\end{tabular}
That is, among the bindings $\nu(l)$ with the same locus $l$ we pick
an equivalence class with a non-canonical virtual binding
$\langle n,\mathtt{inr}\ d',\overline n\rangle$. If such a class does
not exist, we are done. Let $k$ be the memo key of that class. If
$d'$ is $\bot$, so is the whole $\mathsf{canon}\,\nu\, l$. If not,
it is a pair, containing (the meaning of) the code $d$ for the
right-hand-side of the binding, plus its accompanying virtual bindings
$\nu''$. We update the class $k$ so it now contains the canonical
binding $\langle n,\mathtt{inl}\ d,\overline n\rangle$ and merge the result
with $\nu''$. It may happen that $\nu''$ contains a non-canonical
binding for the same locus $l$ and the same memo key $k$. In fact,
this happens in the
|cack2| example above: evaluating |ack 2| produces virtual
bindings that contain |ack 2| again. The operation $\mathsf{merge}$
folds such bindings, with the already 
canonical $\mathtt{inl}\ d$ used as the right-hand-side
for the $\nu''(l)(k)$ binding~-- thus canonicalization may eventually
terminate.
\aside{We need to elaborate the cack2 example, showing how fixpoint is
  found, and how generated names at recursive invocations come out
  fresh.}

The form |let rec locus l in e| is semantically almost the same as
|let locus l in e|, differing only in the extra step of
canonicalization of the virtual bindings produced by |e|. After
canonicalization, the virtual bindings are converted into real
bindings in the same way as for |let locus l in e| 
(only we produce one letrec-expression rather than a nest of
let-expressions, and therefore do not bother with $\mathsf{ordered}$.)
\aside{later on, expand}

\section{MetaOCaml Implementation}
\label{s:metaocaml}

Formal semantics is not the end, but the means; it is developed to be
used. One application, the subject of future work, is reasoning about
generating programs and making sure the generated code is not only
well-formed and well-typed but also intended. Another application, the
subject of the present paper, is to clarify edge cases, and attempt to
minimize them once exposed. The formal development has already proven
useful: it has improved our previous design for generating mutually
recursive definitions \cite{genletrec}, which led to the
straightforward implementation in the current (N111) version of
MetaOCaml \cite{design-and-implementation-ber}. We now briefly
describe the implemented interface.

The let-insertion interface in MetaOCaml is as follows:
\begin{code}
type locus
val locus_global : locus
val genlet : ?name:string -> ?locus:locus -> 'a code -> 'a code
val with_locus : (locus -> 'w code) -> 'w code

type locus_rec
val mkgenlet : ?name:string -> locus_rec -> ('k->'k->bool) ->
  (('k -> ('a->'b) code) -> ('k -> ('a->'b) code))
val with_locus_rec : (locus_rec -> 'w code) -> 'w code
\end{code}
Here |genlet| is a version of |genlet l e_m e| described earlier, for
the case of all memo keys being distinct. The first optional argument
of the MetaOCaml |genlet| is the hint for the variable name to
generate (useful if one wishes to see variable names in the generated
code other than |t_34| and the like.) Locus is the second argument: if
omitted, it defaults to |locus_global|, which is the implicit locus at
the very beginning of the program. The current interface (and also the
implementation) thus subsumes the locus-less |genlet| of the previous
MetaOCaml version.

\sloppy
The type |locus_rec| and the operations |mkgenlet| and 
|with_locus_rec| deal with potentially mutually recursive 
memoizing let-insertion. In the earlier |cack2| we have seen that the
memoization key occurs also in the expression to bind, as an argument
to some function. Such a pattern seems common, and |mkgenlet|
interface is built around it. Furthermore, memo keys are no longer integers;
therefore, the user has to supply the key comparison function
|'k->'k->bool|. Again, it is better to see an example~-- the same
Ackermann function specialization example, but written this time in
MetaOCaml (the example is part of the MetaOCaml test suite).

\fussy
\begin{code}
let sack m =
  with_locus_rec @@ fun l ->
  let g = mkgenlet l (=) in
  let rec loop m =
    if m = 0 then .<fun n -> n + 1>. else
    .<fun n -> if n = 0 then .~(g loop (m-1)) 1
        else .~(g loop (m-1)) (.~(g loop m) (n-1))>.
  in g loop m
\end{code}
The interface is designed so that we could blindly put |g|
before every recursive call, and obtained the desired generator.
The generated code is identical to that for |cack2| shown earlier.

The major difference of MetaOCaml let-insertion is that it never
produces ill-scoped code (that is, code with a scope extrusion).  In
MetaOCaml, let(rec) is actually inserted either at the place of the
explicit |let locus|, or at the binding that dominates all free
variables of the bound expression, whichever has the narrowest scope.

\section{Related Work}
\label{s:related}

It was recognized early
on~\cite{Bondorf:1992:IBT:141471.141483,Lawall:1994:CPE:182409.182483}
that one can use control effects (either direct or realized via CPS)
to answer the compositionality challenge of the ordinary, well-nested
let-insertion. \cite{shifting-the-stage} give a comprehensive formal
treatment.  Unfortunately, neither the standard CPS nor the
well-understood |shift| operator are of any help with let-insertion
that does not follow the stack discipline and crosses
already-generated bindings. This problem was discussed in
\cite{hygienic-combinators}, which proposed a very complicated
transformation for hygienic let-insertion across bindings, whose
correctness was only conjectured. The theory of code movement across
already-generated bindings was later developed in
\cite{DBLP:conf/aplas/KiselyovKS16}, using operational semantics; it
did not include let-insertion however.

Semantics for multi-stage languages have from the very earliest
works~\cite{msp-axiomatization} generally been given operationally;
our denotational presentation is unusual in this respect.  In contrast,
earlier work on two-level languages used a similar style to ours:
\cite{nielson-two-level} give a denotational semantics in which the
meta-language semantics is parameterized by the semantics of the base
language.  (However, that work did not investigate let-insertion.)

Generating (mutually) recursive bindings has not previously been
formally considered at all, to our knowledge.

\section{Conclusions}
We have developed an executable
denotational semantics for let(rec) insertion.  The next step is to
develop a type system that prevents scope extrusion. Our semantics,
for the first time, lets us reason about the code with the generated
let-statements, and we plan to demonstrate this facility on standard
interesting examples
(e.g.~from~\cite{shifting-the-stage,genletrec,DBLP:conf/aplas/KiselyovKS16}).

\section*{Acknowledgments}
We are grateful to the anonymous reviewers for many very helpful
suggestions, which greatly improved the presentation.
We thank Yukiyoshi Kameyama for hospitality.
This work was partially supported by JSPS KAKENHI Grant Number
18H03218.

\bibliographystyle{alpha}
\bibliography{refs}

\begin{thebibliography}{KMS21}

\bibitem[Bon92]{Bondorf:1992:IBT:141471.141483}
Anders Bondorf.
\newblock Improving binding times without explicit {CPS}-conversion.
\newblock In {\em Proceedings of the 1992 ACM Conference on LISP and Functional
  Programming}, LFP '92, pages 1--10, New York, NY, USA, 1992. ACM.

\bibitem[Kis14]{design-and-implementation-ber}
Oleg Kiselyov.
\newblock The design and implementation of {BER} {MetaOCaml}.
\newblock In Michael Codish and Eijiro Sumii, editors, {\em Functional and
  Logic Programming}, volume 8475 of {\em Lecture Notes in Computer Science},
  pages 86--102. Springer International Publishing, 2014.

\bibitem[Kis17]{having-effect}
Oleg Kiselyov.
\newblock Higher-order programming is an effect, 2017.
\newblock HOPE Workshop at ICFP 2017.

\bibitem[KKS11]{shifting-the-stage}
Yukiyoshi Kameyama, Oleg Kiselyov, and Chung-chieh Shan.
\newblock Shifting the stage: Staging with delimited control.
\newblock {\em Journal of Functional Programming}, 21(6):617--662, November
  2011.

\bibitem[KKS15]{hygienic-combinators}
Yukiyoshi Kameyama, Oleg Kiselyov, and Chung-chieh Shan.
\newblock Combinators for impure yet hygienic code generation.
\newblock {\em Science of Computer Programming}, 112 (part 2):120--144,
  November 2015.

\bibitem[KKS16]{DBLP:conf/aplas/KiselyovKS16}
Oleg Kiselyov, Yukiyoshi Kameyama, and Yuto Sudo.
\newblock Refined environment classifiers - type- and scope-safe code
  generation with mutable cells.
\newblock In Atsushi Igarashi, editor, {\em Programming Languages and Systems -
  14th Asian Symposium, {APLAS} 2016, Hanoi, Vietnam, November 21-23, 2016,
  Proceedings}, volume 10017 of {\em Lecture Notes in Computer Science}, pages
  271--291, 2016.

\bibitem[KMS21]{NBEA}
Oleg Kiselyov, Shin-Cheng Mu, and Amr Sabry.
\newblock Not by equations alone: Reasoning with extensible effects.
\newblock {\em Journal of Functional Programming}, 31:e2, 2021.

\bibitem[KS16]{Eff-in-OCaml}
Oleg Kiselyov and K.~C. Sivaramakrishnan.
\newblock Eff directly in {OCaml}.
\newblock In Kenichi Asai and Mark~R. Shinwell, editors, {\em Proceedings {ML}
  Family Workshop / OCaml Users and Developers workshops, {ML/OCAML} 2016,
  Nara, Japan, September 22-23, 2016}, volume 285 of {\em {EPTCS}}, pages
  23--58, 2016.

\bibitem[LD94]{Lawall:1994:CPE:182409.182483}
Julia~L. Lawall and Olivier Danvy.
\newblock Continuation-based partial evaluation.
\newblock In {\em Proceedings of the 1994 ACM Conference on LISP and Functional
  Programming}, LFP '94, pages 227--238, {N}ew {Y}ork, 1994. ACM.

\bibitem[Mos90]{Mosses-denot}
Peter~D. Mosses.
\newblock Denotational semantics.
\newblock In J.~van Leewen, editor, {\em Handbook of Theoretical Computer
  Science}, volume B: Formal Models and Semantics, chapter~11, pages 577--631.
  The MIT Press, New York, NY, 1990.

\bibitem[NN92]{nielson-two-level}
Flemming Nielson and Hanne~Riis Nielson.
\newblock {\em Two-Level Functional Languages}.
\newblock Cambridge University Press, Cambridge, 1992.

\bibitem[ORP16]{RandIR}
Georg Ofenbeck, Tiark Rompf, and Markus P{\"u}schel.
\newblock Rand{IR}: differential testing for embedded compilers.
\newblock In {\em Proceedings of the 7th {ACM} {SIGPLAN} Symposium on Scala,
  {SCALA}@{SPLASH} 2016}, pages 21--30. ACM, October 30 - November 4 2016.

\bibitem[SK01]{sumii-hybrid}
Eijiro Sumii and Naoki Kobayashi.
\newblock A hybrid approach to online and offline partial evaluation.
\newblock {\em Higher-Order and Symbolic Computation}, 14(2--3):101--142,
  September 2001.

\bibitem[TBS98]{msp-axiomatization}
Walid Taha, Zine{-}El{-}Abidine Benaissa, and Tim Sheard.
\newblock Multi-stage programming: Axiomatization and type safety.
\newblock In Kim~Guldstrand Larsen, Sven Skyum, and Glynn Winskel, editors,
  {\em Automata, Languages and Programming, 25th International Colloquium,
  ICALP'98, Aalborg, Denmark, July 13-17, 1998, Proceedings}, volume 1443 of
  {\em Lecture Notes in Computer Science}, pages 918--929. Springer, 1998.

\bibitem[YK19]{genletrec}
Jeremy Yallop and Oleg Kiselyov.
\newblock Generating mutually recursive definitions.
\newblock In {\em Proceedings of the 2019 ACM SIGPLAN Workshop on Partial
  Evaluation and Program Manipulation}, PEPM 2019, pages 75--81, New York, NY,
  USA, 2019. ACM.

\end{thebibliography}

\appendix

\section{Base and Codec, executable}
\label{s:sigs}

The base calculus can be represented as an OCaml signature. Calculus
expressions of type |'a| are represented as OCaml values of
type |'a repr|. The mutually recursive |mletrec| takes a collection
of clauses indexed by |idx|; the first argument to |mletrec| indicates the 
number of clauses. We gave |mletrec| a general type, with polymorphic |'a|.
In all practical cases, |'a| should be a function type and
| ('a->'b) repr| should start with a lambda.
The form |letrec| is an instance of the general |mletrec|.

The |R| and |S| semantics of the calculus are then implementations
of the signature. Using OCaml to express the denotational semantics
was argued in \cite[\S3.1.4]{Eff-in-OCaml}.

\begin{figure}[htbp]
\begin{code}
type 'a repr

val lam  : ('a repr -> 'b repr) -> ('a->'b) repr
val let_ : 'a repr -> ('a repr ->'b repr) -> 'b repr
val (/)  : ('a->'b) repr -> ('a repr -> 'b repr) (* application *)
val if_  : bool repr -> 'a repr -> 'a repr -> 'a repr

type idx = int
val mletrec : idx -> 
  ((idx -> 'a repr) -> (idx -> 'a repr)) ->
  ((idx -> 'a repr) -> 'w repr) -> 'w repr

val letrec : (('a->'b) repr -> 'a repr -> 'b repr) ->
      (('a->'b) repr -> 'w repr) -> 'w repr

val int  : int -> int repr
val bool : bool -> bool repr

val succ   : int repr -> int repr
val ( +  ) : int repr -> int repr -> int repr
val ( -  ) : int repr -> int repr -> int repr
val ( *  ) : int repr -> int repr -> int repr
val ( =. ) : int repr -> int repr -> bool repr
\end{code}
\caption{Base calculus represented in OCaml: its syntax as OCaml signature}
\label{f:base-OCaml}
\end{figure}
\clearpage

\begin{figure}[ht]
\begin{code}
type 'a cd  (* the type of code values *)

val clam  : ('a cd repr -> 'b cd repr) -> ('a->'b) cd repr
val clet : 'a cd repr -> ('a cd repr -> 'b cd repr) -> 'b cd repr

val cint : int  -> int cd repr
val cbool : bool -> bool cd repr

val csucc : int cd repr -> int cd repr

val ( +% ) : int cd repr -> int cd repr -> int cd repr
val ( -% ) : int cd repr -> int cd repr -> int cd repr
val ( *% ) : int cd repr -> int cd repr -> int cd repr
val (=%) : int cd repr -> int cd repr -> bool cd repr

val (/%) : ('a -> 'b) cd repr -> 'a cd repr -> 'b cd repr
val cif_   : bool cd repr -> 'a cd repr -> 'a cd repr -> 'a cd repr

type memo_key = int repr
type locus_t
val genlet_locus : (locus_t -> 'w cd repr) -> 'w cd repr
val genlet : locus_t -> memo_key -> 'a cd repr -> 'a cd repr 

val genletrec_locus : (locus_t -> 'w cd repr) -> 'w cd repr
val genletrec : locus_t -> memo_key -> ('a->'b) cd repr -> ('a->'b) cd repr 
\end{code}
\caption{Codec calculus represented in OCaml: its syntax as OCaml
  signature. The calculus includes the whole Base calculus; only the
  extension is shown.}
\label{f:codec-OCaml}
\end{figure}

Figure~\ref{f:codec-OCaml} presents the |Codec| calculus in the form
of an OCaml signature. To be precise, the figure shows only the
extension of |Base|, with the type of code values and combinators to
produce them. A code value is annotated only with the type of the
expression it generates, with no further classifiers (at present).

The combinator |clam| builds a lambda-expression given the open
expression for the function body. Let-insertion combinators are
the simple, local
let-insertion |clet| demonstrated in \S\ref{s:intro}; 
|genlet| for non-recursive definitions, described in
\S\ref{s:genlet} and |genletrec| for mutually recursive definitions,
\S\ref{s:genletrec}. The locus argument of the latter two is
a variable bound by
the corresponding |genlet_locus|, not an expression; therefore its
type is just |locus_t| rather than |locus_t repr|.
\end{document}